\documentclass[aps,prd,twocolumn,nofootinbib]{revtex4-1}

\usepackage{graphicx}

\newcommand{\tmtexttt}[1]{{\ttfamily{#1}}}

\begin{document}

\title{BCVEGPY2.2: A Newly Upgraded Version for Hadronic Production of the Meson $B_c$ and Its Excited States}

\author{Chao-Hsi Chang}
\email{zhangzx@itp.ac.cn}
\address{State Key Laboratory of Theoretical Physics, Institute of Theoretical Physics, Chinese Academy of Sciences, P.O. Box 2735, Beijing
100080, P.R. China}
\address{CCAST (World Laboratory), P.O. Box 8730, Beijing 100080, China}

\author{Xian-You Wang}
\email{xianyouwang@itp.ac.cn}
\address{State Key Laboratory of Theoretical Physics, Institute of Theoretical Physics, Chinese Academy of Sciences, Beijing 100080, P.R.
China}
\address{National Center for Nanoscience and Technology of China No.11, BeiyitiaoZhongguancun, Beijing 100190, P.R. China}

\author{Xing-Gang Wu}
\email{wuxg@cqu.edu.cn}
\address{Department of Physics, Chongqing University, Chongqing 401331, P.R. China}

\date{\today}

\begin{abstract}
A newly upgraded version of the BCVEGPY, a generator for hadronic production of the meson $B_c$ and its excited states, is available. In comparison with the previous one [C.H. Chang, J.X. Wang and X.G. Wu, Comput. Phys. Commun. {\bf 175}, 624 (2006)], the new version is to apply an improved hit-and-miss technology to generating the un-weighted events much more efficiently under various simulation environments. The codes for production of $2S$-wave $B_c$ states are also given here.
\end{abstract}

\maketitle

\noindent{\bf NEW VERSION PROGRAM SUMMARY} \\

\noindent{\it Title of program} : BCVEGPY2.2 \\

\noindent{\it Program obtained from} : CPC Program Library \\

\noindent{\it Reference to original program} : BCVEGPY2.1 \\

\noindent{\it Reference in CPC}: Comput. Phys. Commun. {\bf 175}, 624 (2006) \\

\noindent{\it Does the new version supersede the old program}: Yes \\

\noindent{\it Computer} : Any LINUX based on PC with FORTRAN 77 or FORTRAN 90 and GNU C compiler as well \\

\noindent{\it Operating systems} : LINUX \\

\noindent{\it Programming language used} : FORTRAN 77/90\\

\noindent{\it Memory required to execute with typical data} : About 2.0 MB \\

\noindent{\it No. of bytes in distributed program} : About 2 MB, including PYTHIA6.4 \\

\noindent{\it Distribution format} : .tar.gz \\

\noindent{\it Nature of physical problem} : Hadronic Production of $B_c$ meson and its excited states. \\

\noindent{\it Method of solution} : To generate un-weighted events of $B_c$ meson and its excited states by using an improved hit-and-miss
technology. \\

\noindent{\it Reasons for new version} : Responding to the feedback from users, such as those from CMS and LHCb groups, we create a new hit-and-miss algorithm for generating the un-weighted events. Furthermore, the relevant codes for generating the $2S$-excited state of $B_c$ meson are added, because the excited state production may be sizable
in the new LHC run. \\

\noindent{\it Typical running time} : It depends on which option is chosen to match PYTHIA when generating the full events and also on which
state of $B_c$ meson, either its ground state or its excited states, is to be generated. Typically on a 2.27GHz Intel Xeon E5520 processor machine, for producing the $B_c$ meson ground state: I) If setting [IDWTUP=3 and \tmtexttt{unwght}=.true.], it shall adopt the new hit-and-miss technology to generate the un-weighted events, and to generate $10^5$ events takes 30 minutes; II) If setting [IDWTUP=3 and \tmtexttt{unwght}=.false.] or [IDWTUP=1 and IGENERATE=0], it shall generate the weighted events, and to generate $10^5$ events takes $2$ minutes only (the fastest way, for theoretical purpose only); III) As a comparison, if setting [IDWTUP=1 and IGENERATE=1], it shall, as the same as the previous version, adopt the PYTHIA inner hit-and-miss technology to generate the un-weighted events, and to generate 1000 events takes about 22 hours. Thus, the efficiency (and accuracy also) for generating the un-weighted events obviously is greatly increased. \\


\noindent{\it Summary of the changes} : 1). We improve the approach for generating un-weighted events. 2). Responding to the feedback from users,
we adjust part of the codes to make it work more user-friendly. More specifically, we explain main changes in the following :

\begin{itemize}

\item {\bf Event generation}.

    If each simulated event comes with a weight, it will make the data analysis much more complicated. Thus the un-weighted events are usually adopted for Monte Carlo simulations. As an external process of PYTHIA, the generator BCVEGPY~\cite{bcvegpy1, bcvegpy2, bcvegpy21, bcvegpy21a} shall call the PYTHIA inner hit-and-miss mechanism to generate the un-weighted events by setting IDWGTUP $=1$ and IGENERATE$=1$~\cite{pythia6}, i.e. the Von Neumann method is used for generating the un-weighted $B_c$ events.

    Every events bearing a weight (\tmtexttt{xwgtup}) respectively, when inputting them to PYTHIA, they are suffered from being accepted or rejected, all the fully generated events at the output become to have a common weight. The Von Neumann method states that the event should be accepted by the PYTHIA subroutine PYEVNT with a probability ${\cal R}$ =\tmtexttt{xwgtup}/\tmtexttt{xmaxup}. This can be achieved by comparing ${\cal R}$ with a random number that is uniformly distributed within the region of $[0,1]$. Namely if ${\cal R}$ is bigger than such a random number then the event is accepted, otherwise it should be rejected. Here \tmtexttt{xmaxup} stands for the maximum event weight.

    The von Neumann method works effectively for the cases when all the weights of input events are moderate in the whole phase-space.
    However if the input events' weights vary greatly, such as varying logarithmically, then its efficiency shall be greatly depressed, since too much time shall be wasted for calculating \tmtexttt{xwgtup} of the rejected events. Thus it is helpful to find a new method for generating un-weighted events.

    We will adopt the new hit-and-miss strategy suggested by Ref.\cite{genxicc21} to do the $B_c$ meson un-weight simulation. Extra switches for calling this new technology are added to BCVEGPY, e.g. the new hit-and-miss technology shall be called by setting IDWTUP=3 and \tmtexttt{unwght}=.true.. Details for this new technology can be found in Ref.\cite{genxicc21}. For self-consistency, we repeat its main idea here.

    To be different from previous versions, BCVEGPY2.2 uses the VEGAS~\cite{vegas} and the MINT~\cite{mint} as a combined way to generate the un-weighted events. The whole phase space shall be separated to a multi-dimensional phase-space grid. The main purpose of VEGAS~\cite{vegas} is to perform the adaptive Monte Carlo multi-dimensional integration, which uses the importance-sampling method to improve the integration efficiency. Each event shall generally result in a different weight, recorded by \tmtexttt{xwgtup}, and the maximum weight within each grid shall be simultaneously recorded into the importance-sampling grid file (with the suffix .grid). Then following the idea of MINT, the Von Neumann method is used in each phase-space grid. Within this small grid region, the von Neumann algorithm works effectively, thus the efficiency for generating un-weighted events are greatly increased.

    To implement the new hit-and-miss algorithm into BCVEGPY2.2, we change the original VEGAS subroutine as
    \begin{widetext}
    \centering
    \tmtexttt{vegas(fxn,ndim,ncall,itmx,nprn,xint,xmax,imode)}
    \end{widetext}
    Three new variables \tmtexttt{xint}, \tmtexttt{xmax} and \tmtexttt{imode} are added in the VEGAS subroutine. The \tmtexttt{xmax} array is used to record the maximum weights in all cells and \tmtexttt{imode} is a flag. \tmtexttt{xint} stands for the output cross-section when setting \tmtexttt{imode=0}, which shall be used to initialize the \tmtexttt{xmax} array when setting \tmtexttt{imode=1}. For convenience, the generated \tmtexttt{xmax} array will be stored in the same grid file in which the importance sampling function is stored.

    In the initialization stage, the VEGAS subroutine shall be called by the subroutine \tmtexttt{evntinit} twice by setting
    \tmtexttt{imode=0} and \tmtexttt{imode=1} respectively to generate both the upper bound grid \tmtexttt{xmax} for all cells and the
    importance sampling function.

    A subroutine \tmtexttt{gen(fxn,ndim,xmax,jmode)} is defined in the file \tmtexttt{vegas.F} with the purpose to generate the un-weighted events. Three options for calling \tmtexttt{gen} subroutine are defined: \tmtexttt{jmode=0} is to initializes the parameter;
    \tmtexttt{jmode=3} is to print the generation statistics; \tmtexttt{jmode=1} is the key option, which is to use the new hit-and-miss technology to generate the un-weighted events. More explicitly, by calling \tmtexttt{gen(fxn,ndim,xmax,jmode=1)}, three steps shall be executed:
    \begin{enumerate}
     \item Call the \tmtexttt{phase$\_$gen} subroutine to generate a random phase-space point and to calculate its weight
         \tmtexttt{xwgtup}.
     \item Judge the point locates in which cell and read from the \tmtexttt{xmax} array and get the upper bound value \tmtexttt{xmaxup} for this particular cell.
      \item Judge whether such point be kept or not by using the Von Neumann method with the help of the probability
          \tmtexttt{xwgtup}/\tmtexttt{xmaxup}.
     \end{enumerate}

    To be more flexible, we add one parameter \tmtexttt{igenmode} for generating or using the existed \tmtexttt{.grid} files. When setting \tmtexttt{igenmode=1}, the VEGAS subroutine shall be called to generate the \tmtexttt{.grid} files. When setting \tmtexttt{igenmode=2}, the VEGAS subroutine shall be called to generate more accurate \tmtexttt{.grid} files from the existed \tmtexttt{.grid} files. When setting \tmtexttt{igenmode=3}, one can directly use the existed \tmtexttt{.grid} files to generate events without running VEGAS. Importantly, before using the existed \tmtexttt{.grid} files, one must ensure all the parameters be the same as the previous generation.

\item {\bf A script for setting the parameters and a cross-check of the un-weighted events.}

    We put an additional file, \tmtexttt{bcvegpy\_set\_par.nam}, in the new version for setting the parameters. This way the user does
    not need to compile the program again if only the parameter values are changed.

    \begin{figure}[tb]
    \centering
    \includegraphics[width=0.45\textwidth]{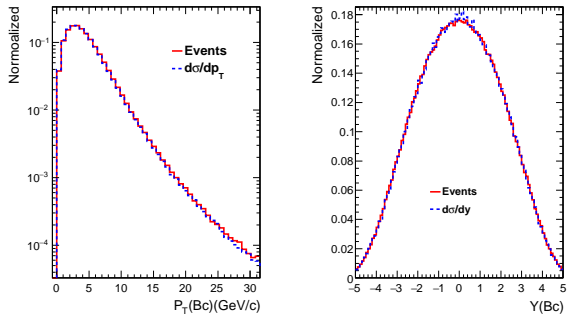}
    \caption{Comparison of the normalized $B_c$ transverse momentum ($P_T$) and rapidity ($y$) distributions derived by setting
    \tmtexttt{unwght}=.true. (events) and \tmtexttt{unwght}=.false. (differential distributions), which are represented by solid and dotted lines, respectively. }
    \label{fig1}
    \end{figure}

    As a cross-check of the new technology, we compare the un-weighted $B_c$ event distributions derived by setting \tmtexttt{unwght}=.true. with the weighted $B_c$ differential distributions derived by setting \tmtexttt{unwght}=.false.. The results are shown in FIG.\ref{fig1}. Those two distributions after proper normalization agree well with each other, that shows our present scheme for un-weighted events is correct.

\item {\bf Bc(2S) generation}.

   In 2014 the ATLAS collaboration reported an observation about an excited state of $B_c$ meson, which most probably is $B_c(2S)$
   state~\cite{Bc2sd}. With more data being collected at LHC detectors, it is hopeful that more observations on the excited $B_c$ states
   will be issued. Therefore in addition to the production via color-singlet $B_c(1S)$, $B_c(1P)$ and color-octet $B_c(1S)$ states,
   the $B_c(2S)$ production is involved in BCVEGPY2.2. It is achieved by replacing the $1S$-wave bound-state parameters
   \tmtexttt{pmb}, \tmtexttt{pmc} and \tmtexttt{fbc} with those of the $2S$-wave one. Here \tmtexttt{pmb}, \tmtexttt{pmc} and \tmtexttt{fbc} are for $b$-quark mass, $c$-quark mass and the radial wave function at the zero ($|R(0)|$), respectively. For the $2S$-wave case, their
   default values are set as \tmtexttt{pmb}=5.234 GeV, \tmtexttt{pmc}=1.633 GeV and \tmtexttt{fbc}=0.991 GeV$^{3/2}$~\cite{fbc} if the mass of the $2S$-wave $B_c$ state is $6.867$ GeV.

   More explicitly, two new values for \tmtexttt{ibcstate} are added: \tmtexttt{ibcstate}=9 is to generate $2^1S_0$ state and
   \tmtexttt{ibcstate}=10 is to generate $2^3S_1$ state. Detailed technologies for deriving the production properties of all the mentioned ten $B_c$ meson states can be found in Refs.\cite{Chang:2003cr, Chang:2004bh, Chang:2005bf}. Furthermore, the values for \tmtexttt{mix\_type} are rearranged. \tmtexttt{mix\_type=1} is to generate the mixing events for all mentioned states. \tmtexttt{mix\_type=2} is to generate the mixing events for $1^1S_0$ and $1^3S_1$ states. \tmtexttt{mix\_type=3} is to generate the mixing events for the four $1P$-wave states and the two color-octet $1^1S_0$ and $1^3S_1$ states. \tmtexttt{mix\_type=4} is to generate the mixing events for $2^1S_0$ and $2^3S_1$ states.

   \end{itemize}

\vspace{1cm}

{\bf Acknowledgments}: This work was supported in part by National Basic Research Program of China (973 program, No.2013CB932804), by the
Natural Science Foundation of China under Grant No.11275243, 11747001 and No.11275280, and by the Fundamental Research Funds for the Central Universities under the Grant No.CDJZR305513.

\end{document}